\pdfoutput=1
\documentclass[pra,twocolumn,amsmath,amssymb,footinbib]{revtex4-1}

\usepackage{graphicx}% Include figure files
\usepackage{dcolumn}% Align table columns on decimal point
\usepackage{bm}% bold math
\usepackage[utf8]{inputenc}
\usepackage{times} % font type
\usepackage[english]{babel}
\usepackage{color}
\begin{document}

%\preprint{APS/123-QED}

\title{Charge- and Flux-Insensitive Tunable Superconducting Qubit}% Force line breaks with \\

\author{Eyob A. Sete}
%\email{eyob@rigetti.com}
\author{Matthew J. Reagor}%
\author{Nicolas Didier}
\author{Chad T. Rigetti}

\affiliation{Rigetti Computing,
 775 Heinz Avenue, Berkeley, California 94710, USA}%

\date{\today}% It is always \today, today,

\begin{abstract}
Superconducting qubits with \textit{in-situ} tunable properties are important for constructing a quantum computer. Qubit tunability, however, often comes at the expense of increased noise sensitivity. Here, we propose a flux-tunable superconducting qubit that minimizes the dephasing due to magnetic flux noise by engineering controllable flux ``sweet spots'' at frequencies of interest. This is realized by using a SQUID with asymmetric Josephson junctions shunted by a superinductor formed from an array of junctions. Taking into account correlated global and local noises, it is possible to improve dephasing time by several orders of magnitude. The proposed qubit can be used to realize fast, high-fidelity two-qubit gates in large-scale quantum processors, a key ingredient for implementing fault-tolerant quantum computers.
\end{abstract}
\pacs{Valid PACS appear here}% PACS, the Physics and Astronomy
\maketitle

%\tableofcontents

\section{Introduction}
Superconducting qubits are promising components for the construction of practical quantum computers. A persistent challenge is two-qubit gate errors, in particular for multiqubit entangling operations. It is desirable to have long coherence times and fast, accurate controls. However, achieving both of these objectives at once has been difficult for superconducting qubits. Some architectures rely on fixed-frequency qubits with static qubit-qubit couplings with two-qubit gates activated by microwaves. While such systems exhibit long coherence times approaching $100~\mu \text{s}$, two-qubit gates typically have a few hundred nanoseconds duration and fidelities up to $F=99.1\%$~\cite{Jerry,Anton,Takita}. Other architectures rely on frequency-tunable qubits, which often have shorter coherence times ($20\textup{--}50~\mu \text{s}$) but faster two-qubit gates ($\sim 50~\text{ns}$), limiting fidelities thus far to $F=99.44\%$ ~\cite{Barends,Kelly}. Fluctuations of qubit frequency due to flux noise significantly affect coherence times ~\cite{martinis14,Quintana17,Kou16}.
Additionally, most tunable-qubit-based entangling gates are implemented using slightly anharmonic qubits, which places limits on control accuracy and results in leakage to upper levels \cite{Chen16}. Achieving flux insensitivity with tunable, highly anharmonic qubits could further bolster superconducting qubit technologies.

In this work, we propose a superconducting qubit design based on a fluxonium~\cite{Vladsc,Koch09,vladthesis,vlad_prb,maslukthesis,kurtisthesis} with engineered flux sweet spots. Its strong nonlinearity, tunability, and insensitivity to noise at its operating frequency makes the qubit a candidate for realizing fast, high-fidelity two-qubit operations. The circuit model of the qubit device consists of a superconducting quantum interference device (SQUID) shunted using an array of Josephson junctions, thus forming two loops \textemdash $A$ and $B$. (see Fig.~\ref{figcircuit}). The qubit's energy spectrum is controlled by the magnetic flux threading through loops A and B. In addition to the sweet spots at the minimum and maximum frequencies, which are already present in the fluxonium, a number of additional sweet spots can be realized when the SQUID has asymmetric junction energies and the magnetic fluxes through loops $A$ and $B$ are related as integer multiples. The number and positions of the sweet spots can be controlled by tuning the asymmetry of the Josephson energies and the area of the loops. Because of its large inductance and relatively small capacitance, the qubit still has a wide tunable frequency range ($\sim~$0.5\textup{--}10 GHz). Our numerical simulations show that when the dephasing is induced by global flux noise (caused by global magnetic field fluctuations), the dephasing time can be improved by several orders of magnitude over fluxonium. However, if the dephasing is induced by independent global noise and local noise (due to spin fluctuations on the surface of the superconductor), the dephasing is limited by the local flux noise. When the two noise sources are correlated, the dephasing time can be improved by several orders of magnitude over fluxonium. This is due to the fact that the two noise sources have opposite sensitivity at a given flux bias point, reducing the dephasing rate.

\section{Model and Hamiltonian}
\begin{figure}[t]
    \centering
    \includegraphics[width=0.9\columnwidth]{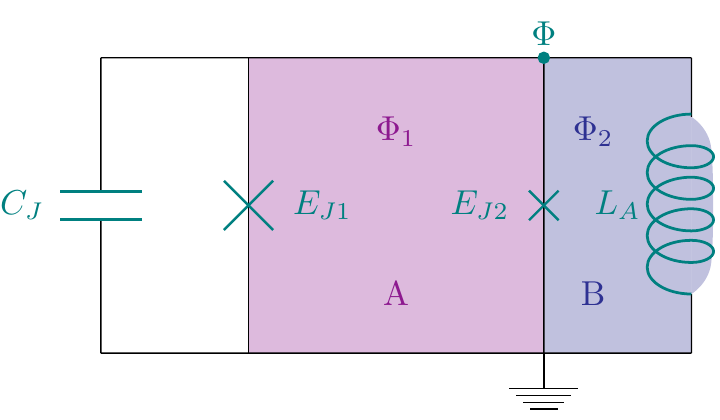}
    \caption{Flatsonium qubit circuit diagram. The circuit consists of a SQUID loop formed by two Josephson junctions of Josephson energies $E_{J1}$, $E_{J2}$ and total capacitance $C_J$. An external magnetic field produces the magnetic flux $\Phi_{1}$ through the SQUID (loop $A$). The SQUID is in turn shunted by an array of Josephson junctions represented by a linear inductor of inductance $L_A$. The area of the loop enclosed by the array inductance $L_A$ and the right SQUID arm is biased by the magnetic flux $\Phi_{2}$ (loop $B$). The three branches yield distinct paths from the node $\Phi$ to the ground node, through the Josephson junctions and the superinductor, and lead to the Hamiltonian equation~\eqref{Hraw}.}
    \label{figcircuit}
\end{figure}

A fluxonium qubit provides broad frequency tunability and large anharmonicity, allowing the construction of fast, high-fidelity two-qubit gates. All flux-tunable qubits are susceptible to flux noise. It has recently been shown that by increasing the asymmetry of the junction energies in a transmon qubit, it is possible to create flux-insensitive points~\cite{Plourde13,Hatch17}, yielding a coherence time that is nearly flux independent~\cite{Hatch17}. In the case of fluxonium, although the $T_1$ time can reach to a few milliseconds when biased at the half-flux quantum~\cite{Pop14}, the dephasing time $T_2$ at the frequency of interest is mostly limited by flux noise~\cite{Pop14,vlad_prb}.
Here, we propose a qubit design that is comprised of a SQUID loop shunted by an array of Josephson junctions (see Fig.~\ref{figcircuit}) and has a flat $|0\rangle\rightarrow |1\rangle$ transition spectrum at the frequency of operation. Hence the choice of the name for our qubit--``flatsonium.''

The proposed circuit consists of a superconducting loop formed by two Josephson junctions of Josephson energies $E_{J1}$ and  $E_{J2}$ with total intrinsic capacitance $C_J$. The superconducting loop $A$ is shunted by an array of Josephson junctions acting as a superinductor of inductance $L_A$. The external flux threading through loop $A$ is $\Phi_{1}$ and the flux through loop $B$ is $\Phi_{2}$. Following the method outlined in Ref.~\cite{Michel97}, the circuit Hamiltonian for flatsonium can be written as
\begin{align}
    H &= 4E_C n^2-E_{J1}\cos \phi-E_{J2} \cos(\phi + 2\pi\Phi_{1}/\Phi_0)\notag\\
    &+\tfrac{1}{2}E_L(\phi + 2\pi\Phi_{1}/\Phi_0 + 2\pi\Phi_{2}/\Phi_0)^2,
    \label{Hraw}
\end{align}
where $n$ is the number of Cooper pairs that have tunneled through the junctions, $\phi=2\pi\Phi/\Phi_0$ is the phase difference between the two nodes ($2\pi$ periodic) with $\Phi_0=h/2e$ being the flux quantum.
%And, $E_{J_{1}}=\Phi_0 I_{c,1}/2\pi=(\Phi_0/2\pi)^2/L_{J_{1}}$ and $E_{J_{2}}=\Phi_0 I_{c,2}/2\pi=(\Phi_0/2\pi)^2/L_{J_{2}}$ are Josephson junction energies,
Similarly, we note $\phi_{j=1,2}=2\pi\Phi_{j}/\Phi_0$.
The Hamiltonian is characterized by the Josephson energies, the charging energy $E_C=e^2/2C_J$ of the Josephson junctions, and the inductance energy $E_L=(\Phi_0/2\pi)^2/L_A$ of the Josephson-junctions array.

We may redefine $\phi$ to move the external flux dependence to the Josephson terms altogether: $\varphi= \phi+\phi_{1}+\phi_{ 2}$ and obtain
\begin{align}\label{q}
    H&=4E_C n^2-E_{J1} \cos(\varphi-\phi_{1}-\phi_{2})
    \notag\\
    &-E_{J2}\cos(\varphi-\phi_{2})+\tfrac{1}{2}E_L\varphi^2.
\end{align}
In the quantum regime, the number $\hat{n}$ of Cooper pairs and the phase $\hat{\varphi}$ are canonical conjugate variables that satisfy the commutation relation $[\hat n, \hat \varphi]= -i$.
Let us assume that the external fluxes through the loops have a general relation  $\Phi_{1} = r \Phi_{2}$, $r \in \mathbb{R}$. Applying a trigonometric relation, the Hamiltonian can be written in the form
\begin{equation}\label{Ham}
    H=4E_C \hat n^2-E_{J_,\rm eff}(\phi_{2})\cos(\hat\varphi-\varphi_0)+\frac{E_L}{2}\hat\varphi^2,
\end{equation}
where the Josephson potential is controlled by the external magnetic fields as follows,
%$\varphi_{\rm ext, 2}=2\pi \Phi_{\rm ext, 2}/\Phi_{0}$,
\begin{equation}\label{EJeff}
    E_{J_,\rm eff}(\phi_{2})=\frac{E_{J\Sigma}}{1+b}\sqrt{1+b^2+2b\cos(r\phi_{2})},
\end{equation}
\begin{equation}
    \varphi_0 = \arctan\left(\frac{b\sin \phi_{2}+\sin[(1+r)\phi_{2}]}{b\cos \phi_{2}+\cos[(1+r)\phi_{2}]}\right).
\end{equation}
We define the Josephson-junction energy asymmetry parameter $b=E_{J2}/E_{J1}$ and $E_{J\Sigma}=E_{J1}+E_{J2}=E_{J1}(1+b)$ the total Josephson-junction energy. The relation between the two fluxes $\Phi_{1} = r \Phi_{2}$ can be achieved in two ways: global flux biasing and on-chip flux biasing. In the case of global flux biasing, a constant magnetic field is applied by placing a large coil beneath  or around the qubit chip. The parameter~$r$ can then be controlled by changing the areas of the loops. In the case of on-chip flux biasing, a magnetic field is applied locally through an on-chip current-carrying wire. Since the magnetic fluxes through the two loops are variable, the parameter~$r$ is controlled by the applied magnetic field and the area of the loops.

\begin{figure}[t]
    \centering
    \includegraphics[width=0.9\columnwidth]{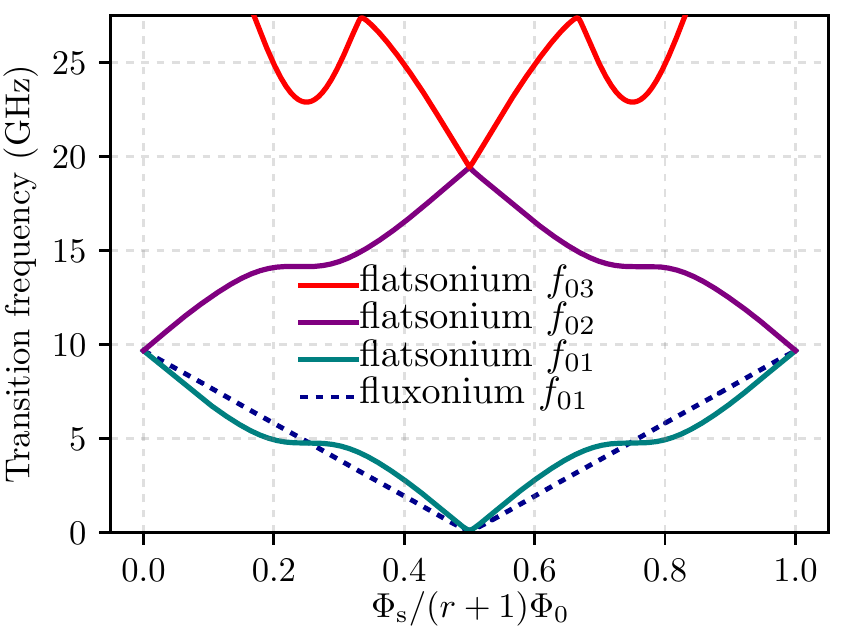}
    \caption{The first three transition frequencies of the flatsonium qubit for
     $E_C/h=6$~GHz, $E_L/h=0.5$~GHz, $E_{J\Sigma}/h=(b+1) E_{J1}/h=20$~GHz, for flux asymmetry $r=2$ (which corresponds to junction energy asymmetry $b=r+1=3$).}
    \label{figT1}
\end{figure}

The fluxonium Hamiltonian can be recovered from Eq.~\eqref{Ham} by setting $b=0$ and $r=0$:
\begin{align}
    H_F=4 E_C\hat n^2-E_{J\Sigma}\cos(\hat\varphi-\phi_{2})+\frac{E_L}{2}\hat\varphi^2,
\end{align}
with the effective junction energy $E_{J\Sigma}$. This Hamiltonian has been shown to have a strongly anharmonic energy spectrum~\cite{vladthesis}. The $|0\rangle\rightarrow |1\rangle$  transition frequency $f_{01}$ has a wide tunability as a function of external fluxes and has flux sweet spots at $\phi_{2}=m\pi$, where $m\in\mathbb{Z}$. Away from the sweet spots, the frequency has a linear dependence on the applied magnetic flux (dashed curve in Fig.~\ref{figT1}) and is approximately given by $f_{01}\approx 4\pi^2E_L E_{J\Sigma}|1/2-\phi_{2}/2\pi|/h(E_L+E_{J\Sigma})$~\cite{vladthesis}. Note that the fluxonium qubit energy parameters broadly fall in the range: $E_L\ll E_C\ll E_{J\Sigma}$ with $2\lesssim E_{J\Sigma}/E_C\lesssim 5$~\cite{vladthesis, kurtisthesis, maslukthesis}. While the energy relaxation time $T_1$ of a fluxonium qubit at operating frequency ($4\textup{--}8$~GHz) is comparable to or better than that of transmon qubits~\cite{inhouse,vlad_prb}, the dephasing time is still very sensitive to flux noise and is limited to a few microseconds~\cite{vlad_prb}.

In addition to the flux sweet spots at zero-flux, half-flux, and one-flux quantum values, others can be created by introducing asymmetry in the SQUID junction energies and the external magnetic fluxes. It turns out that for the circuit shown in Fig.~\ref{figcircuit}, periodic flux-insensitive points and a symmetric energy spectrum can be obtained when the asymmetry parameter $b$ of the two junction energies satisfies $b=r+1$, where $r$ and $b$ are integer values. Note that flux sweet spots can also be created for noninteger $r$ values. However, the resulting spectrum is not symmetric about the half-flux quantum bias point.

In the $f_{01}$ transition frequency, the flux sweet spots appear at the minima of the effective potential $U(\hat\varphi)=E_L\hat\varphi^2/2-E_{J,\rm eff}\cos(\hat \varphi-\varphi_0)$. Setting the first derivative of the potential to zero,  $\partial U(\varphi_0)/\partial \phi_{2}=0$, we find the condition for the flux-insensitive points to be $\phi_{2}/2\pi=m/2r$, where $m\in\mathbb{Z}$. The total number of flux sweet spots $n_{\rm sp}$ within the normalized one-flux quantum is given by (for $r>1$)
\begin{equation}\label{np}
   n_{\rm sp}=\begin{cases}
   r+4, \quad\text{for odd}\ r,\\
   r+3, \quad\text{for even}\ r,
   \end{cases}
\end{equation}
where we have included the flux sweet spots at zero- and at one-flux quantum ($m=0,1,\dots,2r$).

In order to discuss the notion of flux-insensitive points, it is more convenient to introduce the common mode $\Phi_{s}=\Phi_1+\Phi_2$ and the differential mode $\Phi_{d}= \Phi_1-\Phi_2$. In view of the relation $\Phi_{1}=r\Phi_{2}$, the common and differential mode variables are related as $\Phi_{d}= \beta\Phi_{s}$, where $\beta=(r-1)/(r+1)$.

The energy spectrum of the flatsonium qubit can be obtained by numerically diagonalizing the Hamiltonian equation~\eqref{Ham}. Because of the bosonic commutation rule between $\hat{n}$ and $\hat{\varphi}$, it is more convenient to use the Fock basis by writing these conjugate operators as the two quadratures of a harmonic oscillator that is rendered nonlinear by the Josephson junctions.
%(when $E_{J\Sigma}=0$).
Using the first 50 Fock states, energy parameters $E_L/h=0.5$~GHz, $E_C/h=6$~GHz,  $E_{J\Sigma}/h=20$~GHz, loop area asymmetry $r=2$ $(\beta=1/3)$, and junction symmetry $b=r+1=3$, the results of the numerical diagonalization are shown in Fig.~\ref{figT1}. The dashed line represents the fluxonium $f_{01}$ frequency obtained by setting $r=0$ and  $b=0$. The rest of the curves represent the first three transition frequencies of the flatsonium. The $f_{01}$ transition frequency (cyan curve) has two additional flux sweet spots within a normalized one-flux quantum, as expected from Eq.~\eqref{np}. The flux sweet spots appear on either side of the minimum frequency at $\Phi_{s}/(r+1)\Phi_{0}=1/4, 3/4$. Note also that the anharmonic energy spectrum feature is still intact, making the device a good approximation of a two-level system.

\section{Dephasing due to Magnetic flux noise}
In general, dephasing is understood as the fluctuation of the qubit frequency due to its coupling with the environment. Low-frequency noise far below the transition frequency can cause the qubit to accumulate a random phase. It is well known that flux-tunable qubits such as flux and fluxonium qubits are susceptible to flux noise. This is due to the fact that fluctuations in flux induce fluctuations in the qubit frequency, resulting in dephasing. This leads to a short dephasing $T_2$ time. The sensitivity of the qubit frequency to flux noise can be suppressed by engineering flux sweet spots. For an external flux bias, these fluctuations can be induced by a global flux noise (caused by a fluctuating magnetic field) and/or a local flux noise, for example, caused by fluctuating spins at the surface of the superconductor~\cite{Koch07} or in defects at the metal-insulator interface~\cite{Faoro08}. It has been shown for a device with similar circuit topology that the differential mode can be sensitive to local noise and global noise~\cite{Kou16,Nakamura10}.

In the case of weak fluctuations, each external parameter in the qubit Hamiltonian can be decomposed into its controlled dc value and fluctuations around it.
For the external flux, we can write $\Phi_{s}=\bar{\Phi}_{s}+\delta \Phi_{s}$ and $\Phi_{d}=\bar{\Phi}_{d}+
\delta \Phi_{d}$. The Hamiltonian of the qubit, expressed in terms of Pauli operators, can be expanded in a Taylor series as
\begin{align}\label{H-approx}
    H=\frac{\hbar}{2}\left[\omega_{01}+\frac{\partial \omega_{01}}{\partial \Phi_{s}}\delta\Phi_{s}+ \frac{\partial \omega_{01}}{\partial \Phi_{d}}\delta\Phi_{d}+\ldots\right]\hat{\sigma}_z.
\end{align}
The fluctuation generally results in two distinct effects.
For sufficiently low frequencies, the fluctuations cause random shifts in the transition frequency of the qubit, leading to pure dephasing ($T_2$). For higher frequencies, the fluctuations induce transitions between the qubit states, leading to energy relaxation ($T_1$).
In the following, we focus on reducing the pure dephasing rate to improve $T_2$.

The degradation of $T_2$ due to fluctuations in qubit frequency can be understood from the evolution of the off-diagonal element of the qubit density matrix. Considering only the dephasing caused by the magnetic flux noise, the off-diagonal density matrix element can be written as $\rho_{01}(t)=e^{i\omega_{01} t}\langle e^{i\delta v(t)}\rangle$, where the random phase noise is given by
\begin{align}
    \delta v(t)=\frac{\partial \omega_{01}}{\partial \Phi_{s}}\int_0^tdt'\delta\Phi_{s}(t')+\frac{\partial \omega_{01}}{\partial \Phi_{d}}\int_0^tdt'\delta\Phi_{d}(t').
\end{align}
In general, the two noise sources can be correlated \cite{Nakamura10,Gus11}. Correlated flux noise in two inductively coupled flux qubits has been experimentally demonstrated \cite{Nakamura10}. For correlated flux noises and assuming Gaussian noise,
$\langle e^{i\delta v(t)}\rangle=e^{-\frac{1}{2}\langle \delta v^2(t)\rangle}$, the random phase noise has the form (see the Appendix)
\begin{align}
\langle \delta v^2(t)\rangle
&=\left(\frac{\partial \omega_{01}}{\partial \Phi_{s}}\right)^2\int_{-\infty}^{\infty}df\, \text{sinc}^2(\pi f t) S_{\Phi_{s}}(f)\notag\\
&+ \left(\frac{\partial \omega_{01}}{\partial \Phi_{d}}\right)^2\int_{-\infty}^{\infty}df\, \text{sinc}^2(\pi f t) S_{\Phi_{d}}(f)\notag\\
&+2\frac{\partial \omega_{01}}{\partial \Phi_{s}} \frac{\partial \omega_{01}}{\partial \Phi_{d}} \int_{-\infty}^{\infty}df\, \text{sinc}^2(\pi f t) S_{\Phi_s\Phi_{d}}(f),
\end{align}
where $S_{\Phi_s}(f)$ and $S_{\Phi_{d}}(f)$ represent the spectral densities of the global and local noises while $S_{\Phi_s\Phi_{d}}(f)$ represents the spectral density of the correlated noise.  In flux-tunable devices, flux noise is mostly dominated by low-frequency noise and the spectral density has  $1/f$ spectrum~\cite{Ithier05,Koch07}
\begin{equation}
    S_{\Phi}(f)= A_{\Phi}^2/|f|^{\alpha},
\end{equation}
where $0.8\lesssim \alpha \lesssim 1.3$. The parameter $A_{\Phi}$ determines the overall amplitude of the fluctuations and has been measured in various experiments~\cite{Zorin,Well,van,Yoshi,Barends,Omalley15,Ithier05} with values ranging from $A_{\Phi}=10^{-6}\Phi_0$ to $A_{\Phi}=10^{-5}\Phi_0$. The total dephasing rate due to flux noise has the form (see the Appendix)
\begin{align}\label{d-rate}
    \Gamma_{\phi} &= \Big(\Gamma_{\phi, s}^2+\Gamma_{\phi,d}^2\notag\\
    &+2 c_{sd} A_{\Phi_s}A_{\Phi_d}
     \frac{\partial \omega_{01}}{\partial \Phi_d}\frac{\partial \omega_{01}}{\partial \Phi_s}|\ln (\zeta/ f_{\rm ir}t_m)|\Big)^{1/2},
\end{align}
where $\zeta=e^{3/2-\gamma}/2\pi$ with $\gamma$ being the Euler's constant, $t_{m}\sim1/\Gamma_{\phi}$, $c_{sd} = A_{\Phi_s \Phi_d}^2/A_{\Phi_s}A_{\Phi_d}$ is the correlation coefficient, $A_{\Phi_s \Phi_d}$ is the noise amplitude of the correlated noise, and the dephasing rates due to the common-mode and differential-mode noises are, respectively,
\begin{align}
    \Gamma_{\phi,s}&=2\pi A_{\Phi_{s}}\sqrt{|\ln (\zeta/ f_{\rm ir}t_m)|}\left|\frac{\partial f_{01}}{\partial \Phi_{s}}\right|,\label{sum-rate}\\
    \Gamma_{\phi,d}&=2\pi A_{\Phi_{d}}\sqrt{|\ln (\zeta/ f_{\rm ir}t_m)|}\left|\frac{\partial f_{01}}{\partial \Phi_{d}}\right|,
\end{align}
where $\sqrt{|\ln (\zeta/ f_{\rm ir}t_m)|}\sim 4$~\cite{Kou16,Hatch17}, where $f_{\rm ir}$ is the infrared cut-off frequency. With the general relation between the two fluxes defined earlier, $\Phi_{d}=\beta \Phi_{s}$, we can rewrite the dephasing rate due to the local noise as $ \Gamma_{\phi, d}= 2\pi A_{\Phi_{d}}[\sqrt{|\ln (\zeta/ f_{\rm ir}t_m)}/\Phi_{s}]\left|\partial f_{01}/\partial \beta\right|$.
Note that if the two noise sources are uncorrelated the dephasing rate becomes
\begin{align}\label{Gamma-uncorrelated}
\Gamma_{\phi}= \sqrt{\Gamma_{\phi,s}^2+\Gamma_{\phi,d}^2}.
\end{align}
For perfectly correlated noise sources ($c_{sd}=1$) and when the product of the sensitivities $(\partial \omega_{01}/\partial \Phi_s)(\partial \omega_{01}/\partial \Phi_{d})$ is negative, the correlation can lead to a smaller dephasing rate. If they are equal in magnitude, it can effectively cancel the dephasing rate \cite{Gus11}.
\begin{figure}
\includegraphics[width=0.9\columnwidth]{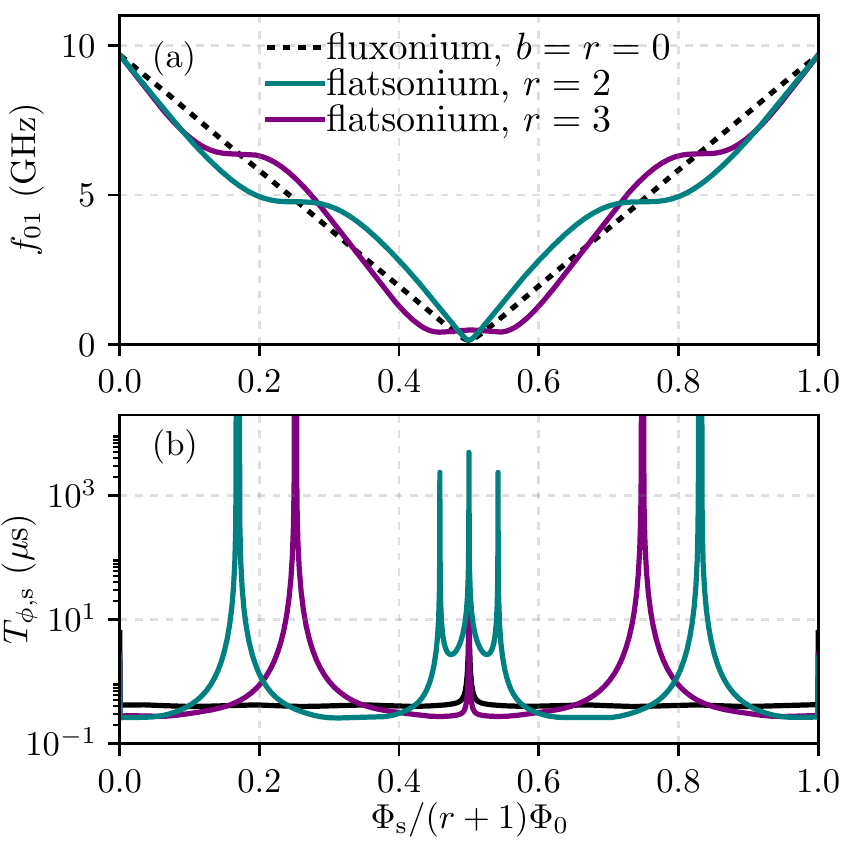}
\caption{(a)~The $|0\rangle\rightarrow|1\rangle$ transition frequency of the fluxonium qubit (dotted) and the flatsonium qubit ($r=2$ in cyan, $r=3$ in magenta) as a function of the total external flux threading the loops.
Parameters are $E_{J\Sigma}/h=20$~GHz, $E_L/h=0.5$~GHz, and $E_C/h=6$~GHz.
(b)~Dephasing time $T_{\phi, s}=1/\Gamma_{\phi, s}$ [Eq.~\eqref{sum-rate}] for the fluxonium and flatsonium qubits. Here, we have used $A_{\Phi_{s}}=5~\mu  \Phi_{0}$ for the common-mode noise and no differential-mode noise, $A_{\Phi_{d}}=0$.}
\label{figT2}

\end{figure}
\begin{figure}[t]
\includegraphics[width=0.9\columnwidth]{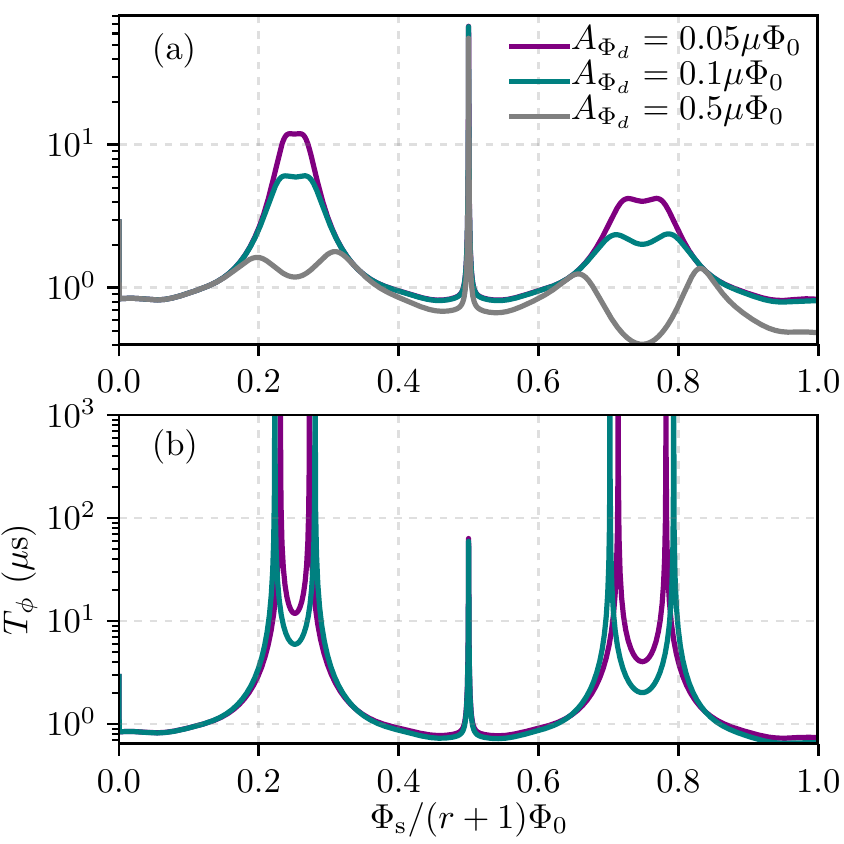}
\caption{ Total dephasing time ($T_{\phi}=1/\Gamma_{\phi})$ versus total flux threading through the two loops with $r=2$ and for various value of local noise amplitude: (a) for uncorrelated noises [Eq. \eqref{Gamma-uncorrelated}] and (b) perfectly correlated noise ($c_{sd}= 1$) [Eq. \eqref{d-rate}]. We used the global noise amplitude $A_{\Phi_{s}}=5~\mu \Phi_{0}$ and all other parameters are the same as in Fig.~\ref{figT2}.}
\label{local_noise}
\end{figure}

%The total dephasing rate due to flux noise is related to $T_2$ as
%\begin{align}\label{dephase}
%       \frac{1}{T_2}=\frac{1}{2T_1}+\Gamma_{\phi}+\Gamma_{\rm other}.
%\end{align}
%At the flux sweet spots, $T_2$ approaches the limit set by the energy relaxation $T_1$ and other dephasing rates, $\Gamma_{\rm other}$ due to, for example, charge noise, critical current noise, etc~\cite{Kou16}.

The dephasing time $T_{\phi, s}=1/\Gamma_{\phi,s}$ induced by qubit-frequency fluctuations due to global flux noise for the flatsonium and the fluxonium qubits is shown in Fig.~\ref{figT2} for realistic parameters~\cite{vlad_prb}. For the fluxonium, due to the linear dependence on the externally applied flux, the dephasing time is constant and limited to $\sim 1 ~\mu\text{s}$ away from the sweet spot at $\Phi_{s}(r+1)/\Phi_{0}=1/2$. If we let the SQUID junctions be asymmetric such that $b=3$ and the asymmetry of fluxes $r=2$, we get two additional flux-insensitive points for the flatsonium (at $\Phi_{s}/(r+1)\Phi_{0}=1/4, 3/4$) as illustrated in Fig.~\ref{figT2}(a) (cyan curve). The corresponding dephasing time [Fig.~\ref{figT2}(b)] at these flux insensitive points is  $T_{\phi}\sim 10~\text{ms}$, an improvement by several orders of magnitude. Increasing the junction asymmetry to $b=4$ ($r=3$), two more flux sweet spots close to the half-flux quantum point are created (purple curve). This is consistent with our prediction in Eq.~\eqref{np}. Note that the position of the other two flux sweet spots move away from the half-flux quantum point and appear at a higher frequency. The dephasing time is still more than $1$~ms.

In Fig.~\ref{local_noise}(a), we show the contribution of both global noise and local noise to the dephasing time for different values of the local noise amplitude $A_{\Phi_{d}}$, assuming that the two noise sources are uncorrelated. Although the qubit is insensitive to global noise at integer multiples of a half-flux quantum, the dephasing at the engineered flux sweet spots is limited by local noise. When the two noise sources are perfectly correlated ($c_{sd}=1$), there exist flux bias points where the contributions of the global and local noises cancel out, giving rise to sweet spots. Figure \ref{local_noise}(b) shows a several-order-of-magnitude improvement in dephasing time at slightly shifted flux biases from the original sweet-spot flux biases. These flux biases correspond to the inflection points in the flatsonium spectrum (see cyan curve in Fig.~\ref{figT2}). Local noise due to spin fluctuations remains one of the limiting factors of the coherence time in flux-tunable qubits. There has been a recent effort \cite{kumar16} to minimize the local noise by implementing surface passivation and improvements in the sample vacuum environment. This approach has led to significant reductions in the surface spin susceptibility and low-frequency flux noise power~\cite{kumar16}.

\section{Conclusion}
We have analyzed a tunable superconducting qubit design with engineered flux sweet spots, which we call a flatsonium. The qubit circuit consists of an asymmetric SQUID shunted by an array of Josephson junctions. The qubit is tuned by an external magnetic flux threaded through the SQUID and the Josephson-junction array loops. With the appropriate choice of Josephson junctions and loop parameters, the low-energy spectrum of the flatsonium exhibits a wide frequency tunability, and flux-insensitive points in the range of frequencies of interest. The SQUID Josephson-junction asymmetry and the ratio of the magnetic fluxes threading the loops determine the number and position of the flux sweet spots.
Assuming that the main contribution to the low-frequency noise comes from correlated global and local noises, the dephasing time can be improved by several orders of magnitude over a corresponding fluxonium qubit at the same flux bias point.
Its tunability, strong anharmonicity, and charge-noise insensitivity~\cite{Vladsc,Koch09} (due to its large inductance), together with the engineered flux sweet spots, make the flatsonium qubit a highly coherent device and a potential candidate for constructing fast, high-fidelity quantum logic gates.\\

\begin{acknowledgements}
The authors thank P. Karalekas, M. Curtis, and W. O'Brien for reading of the manuscript and the rest of the Rigetti Computing team for useful discussions. We also thank Michel Devoret and David Schuster for helpful discussions.
\end{acknowledgements}

\begin{appendix}
\renewcommand{\theequation}{A\arabic{equation}}
\section*{Appendix: Derivation of dephasing rate due to $1/f$ noise}
The frequency of the flatsonium qubit can be varied by applying magnetic flux through the two loops forming the flatsonium circuit. The flux tunability brings unwanted dephasing channels to the qubit, limiting coherence times. Any fluctuations of the flux bias induces dephasing to the qubits. For the circuit considered here, the fluctuation of the global magnetic field can cause dephasing. The other mechanism that can induce (local) noise is fluctuating spins at the surface of the conductor \cite{Faoro08} or in defects at the metal-insulator interface \cite{Zorin}.

Let the total magnetic flux threading the two loops be $\Phi_s=\Phi_1+\Phi_2$  and their difference be $\Phi_d=\Phi_1-\Phi_2$. These fluxes can be written as the sum of their mean values plus small fluctuations: $\Phi_s=\bar\Phi_s+\delta\Phi_s$ and $\Phi_d=\bar\Phi_d+\delta\Phi_d$. These small fluctuations cause the qubit frequency to vary, inducing dephasing to the qubit. Note that the mean values of the fluctuations are zero, i.e., $\langle \delta \Phi_s(t)\rangle=\langle \delta \Phi_d(t)\rangle=0$ and satisfy the following time-average correlation functions \cite{martinis03,Nakamura10,Gus11}:
\begin{align}
&\langle \delta \Phi_s(t)\delta\Phi_s(t+\tau)\rangle =\int_{-\infty}^{\infty}d f S_{\Phi_s}(f)~e^{i 2\pi f\tau},\label{noise1}\\
&\langle \delta \Phi_d(t)\delta\Phi_d(t+\tau)\rangle =\int_{-\infty}^{\infty}d f S_{\Phi_d}(f)~e^{i 2\pi f\tau},\label{noise2}\\
&\langle \delta \Phi_s(t)\delta\Phi_d(t+\tau)\rangle =\int_{-\infty}^{\infty}d f S_{\rm \Phi_s,\Phi_d}(f)~e^{i 2\pi f\tau},\label{noise3}
\end{align}
where the spectral densities have $1/f$ spectrum~\cite{Ithier05,Koch07}:
\begin{align}\label{1overf}
S_{\Phi_s}= A_{\Phi_s}^2/|f|^\alpha,~~~ S_{\Phi_d}=A_{\Phi_d}^2/|f|^\alpha
\end{align}
and cross-correlation spectral density has the form
\begin{align}
S_{\rm \Phi_s,\Phi_d}(f)=A_{\Phi_s \Phi_d}^2/|f|^\alpha
\end{align}
which vanishes for uncorrelated noise sources. Here, $0.8\lesssim \alpha\lesssim 1.3$, and $A_{\Phi_s}$ and $A_{\Phi_d}$ are the noise amplitudes for the global flux and local flux noise, and $A_{\Phi_s \Phi_d}$ is the noise amplitude of the cross-correlation between these two noise sources \cite{Nakamura10,Gus11}.

The low-laying energy of the qubit can then be described as
\begin{equation}
    H = \hbar\left(\omega_{01}+\frac{\partial \omega_{01}}{\partial \Phi_s}\delta\Phi_s(t)+\frac{\partial \omega_{01}}{\partial \Phi_d}\delta\Phi_d(t)+\hdots \right)\frac{\sigma_z}{2}.
\end{equation}
To understand how the fluctuation in frequency translates into dephasing, it is instructive to consider the evolution of the off-diagonal density matrix element
\begin{equation}
    \rho_{01}(t) = e^{i\omega_{01}t}\langle e^{i \delta v(t)}\rangle,
\end{equation}
where
\begin{equation}
    \delta v(t)= \frac{\partial \omega_{01}}{\partial \Phi_s}\int_{0}^tdt'\delta\Phi_s(t')+\frac{\partial \omega_{01}}{\partial \Phi_d}\int_{0}^{t}dt'\delta\Phi_d(t')
\end{equation}
is the phase noise of the qubit state due to the random rotations. %Expanding the exponential $e^{i v(t)}$, we get
%\begin{equation}
%     \langle e^{i v(t)}\rangle = 1 + i\langle \delta v(t)\rangle -\frac{1}{2}\langle  \delta v(t)^2\rangle-i\frac{1}{2}\langle  \delta v(t)^3\rangle+ \frac{1}{4!}\langle  \delta v(t)^4\rangle+\hdots
%\end{equation}
For a Gaussian noise and for the vanishing mean of the random phase noise $\langle v(t)\rangle=0$, we have
\begin{align}
 \langle e^{i v(t)}\rangle=e^{-(1/2)\langle \delta v^2(t)\rangle}.
\end{align}
Therefore, the mean-squared phase noise $\langle \delta v^2(t)\rangle$ can be expressed as \citep{martinis03}
\begin{align}\label{phasenoise}
\langle \delta v^2(t)\rangle &= \left(\frac{\partial \omega_{01}}{\partial \Phi_s}\right)^2\left\langle\int_{0}^{t}dt'\int_{0}^{t}dt''\delta\Phi_s(t')\delta\Phi_s(t'')\right\rangle\notag\\
&+\left(\frac{\partial \omega_{01}}{\partial \Phi_d}\right)^2\left\langle\int_{0}^{t}dt'\int_{0}^{t}dt''\delta\Phi_d(t') \delta\Phi_d(t'')\right\rangle\notag\\
&+\frac{\partial \omega_{01}}{\partial \Phi_s}\frac{\partial \omega_{01}}{\partial \Phi_d}\left\langle\int_{0}^{t}dt'\int_{0}^{t}dt''\delta\Phi_s(t')\delta\Phi_d(t'')\right\rangle\notag\\
&+\frac{\partial \omega_{01}}{\partial \Phi_s}\frac{\partial \omega_{01}}{\partial \Phi_d}\left\langle\int_{0}^{t}dt'\int_{0}^{t}dt''\delta\Phi_d(t')\delta\Phi_s(t'')\right\rangle.
\end{align}
Considering the first term in \eqref{phasenoise}, and substituting Eqs. \eqref{noise1} and \eqref{1overf} [with $\alpha=1$], we have
\begin{align}\label{global}
  &\left(\frac{\partial \omega_{01}}{\partial \Phi_s}\right)^2\left\langle\int_{0}^{t}dt'\int_{0}^{t}dt''\delta\Phi_s(t')\delta\Phi_s(t'')\right\rangle\notag\\
  &= \left(\frac{\partial \omega_{01}}{\partial \Phi_s}\right)^2 \int_{-\infty}^{\infty} df S_{\Phi_s}(f)\int_{0}^tdt'\int_{0}^tdt''~e^{i\omega(t''-t')}\notag\\
  &=2 \left(\frac{\partial \omega_{01}}{\partial \Phi_s}\right)^2 \int_{f_{\rm ir}}^{\infty} d f S_{\Phi_s}(f)\frac{\sin^2(\omega t/2)}{(\omega/2)^2},
\end{align}
where we have introduced infrared cutoff frequency $f_{\rm ir}$, which is determined by the inverse of the total length of the experiment. Substituting the first equation in Eq. \eqref{1overf} into \eqref{global} and performing the resulting integration, we get
\begin{align}\label{gamma1}
&\left(\frac{\partial \omega_{01}}{\partial \Phi_s}\right)^2\left\langle\int_{0}^{t}dt'\int_{0}^{t}dt''\delta\Phi_s(t')\delta\Phi_s(t'')\right\rangle\notag\\
&\simeq 2 \left(\frac{\partial \omega_{01}}{\partial \Phi_s}\right)^2 A_{\Phi_s}^2t^2|\ln (\zeta / f_{\rm ir}t_m)|,
\end{align}
where $\zeta=\exp(3/2-\gamma)/2\pi\approx 0.400479$ with $\gamma$ being the Euler's constant and $t_m\sim 1/\Gamma_{\phi}$. Similarly, the last three terms in Eq. \eqref{phasenoise} yield
\begin{align}
&\left(\frac{\partial \omega_{01}}{\partial \Phi_d}\right)^2\left\langle\int_{0}^{t}dt'\int_{0}^{t}dt''\delta\Phi_d(t') \delta\Phi_d(t'')\right\rangle\notag\\
&\simeq 2\left(\frac{\partial \omega_{01}}{\partial \Phi_d}\right)^2 A_{\Phi_d}^2t^2|\ln (\zeta/ f_{\rm ir}t_m)|\\
&\frac{\partial \omega_{01}}{\partial \Phi_s}\frac{\partial \omega_{01}}{\partial \Phi_d}\left\langle\int_{0}^{t}dt'\int_{0}^{t}dt''\delta\Phi_s(t')\delta\Phi_d(t'')\right\rangle\notag\\
&\simeq 2\frac{\partial \omega_{01}}{\partial \Phi_s}\frac{\partial \omega_{01}}{\partial \Phi_d} A_{\Phi_s \Phi_d}^2t^2|\ln (\zeta/ f_{\rm ir}t_m)|\\
&\frac{\partial \omega_{01}}{\partial \Phi_s}\frac{\partial \omega_{01}}{\partial \Phi_d}\left\langle\int_{0}^{t}dt'\int_{0}^{t}dt''\delta\Phi_d(t')\delta\Phi_s(t'')\right\rangle\notag\\
&\simeq 2\frac{\partial \omega_{01}}{\partial \Phi_d}\frac{\partial \omega_{01}}{\partial \Phi_s} A_{\Phi_s \Phi_d}^2t^2|\ln (\zeta/ f_{\rm ir}t_m)|.
\end{align}
Using these results and Eq. \eqref{gamma1}, the mean-squared phase noise takes the form
\begin{align}
\langle \delta v(t)^2\rangle & = 2\Big[A_{\Phi_s}^2\left(\frac{\partial \omega_{01}}{\partial \Phi_s}\right)^2+A_{\Phi_d}^2\left(\frac{\partial \omega_{01}}{\partial \Phi_d}\right)^2\notag\\
&+2A_{\Phi_s \Phi_d}^2 \frac{\partial \omega_{01}}{\partial \Phi_d}\frac{\partial \omega_{01}}{\partial \Phi_s}\Big] t^2 |\ln (\zeta/ f_{\rm ir}t_m)|.
\end{align}
Therefore, the off-diagonal density matrix element decays as
\begin{align}
\rho_{01}=e^{i\omega_{01}t}e^{-\Gamma_{\phi}^2t^2},
\end{align}
where the dephasing rate $\Gamma_{\phi}$ is given by
\begin{align}
\Gamma_{\phi}&=\sqrt{|\ln (\zeta/ f_{\rm ir}t_m)|}\Big[A_{\Phi_s}^2\left(\frac{\partial \omega_{01}}{\partial \Phi_s}\right)^2+A_{\Phi_d}^2\left(\frac{\partial \omega_{01}}{\partial \Phi_d}\right)^2\notag\\
&+2A_{\Phi_s \Phi_d}^2 \frac{\partial \omega_{01}}{\partial \Phi_d}\frac{\partial \omega_{01}}{\partial \Phi_s}\Big]^{1/2}\notag\\
&=\left(\Gamma_{\Phi_s}^2+\Gamma_{\Phi_d}^2+2A_{\Phi_s \Phi_d}^2 \frac{\partial \omega_{01}}{\partial \Phi_d}\frac{\partial \omega_{01}}{\partial \Phi_s}|\ln (\zeta/ f_{\rm ir}t_m)|\right)^{1/2},
\end{align}
where the contributions of the global and local flux noises to the total dephasing rate are, respectively,
\begin{align}
&\Gamma_{\Phi_{s}}=A_{\Phi_s}\left|\frac{\partial \omega_{01}}{\partial \Phi_s}\right|\sqrt{|\ln (\zeta/ f_{\rm ir}t_m)|},\\
&\Gamma_{\Phi_{d}}=A_{\Phi_d}\left|\frac{\partial \omega_{01}}{\partial \Phi_d}\right|\sqrt{|\ln (\zeta/ f_{\rm ir}t_m)|}.
\end{align}
Introducing correlation coefficient as \cite{Gus11}
\begin{align}
c_{sd} =A_{\Phi_s\Phi_d}^2/A_{\Phi_s}A_{\Phi_d},
\end{align}
the total dephasing rate can be written as
\begin{align}
\Gamma_{\phi}=\left(\Gamma_{\Phi_s}^2+\Gamma_{\Phi_d}^2+2c_{sd}A_{\Phi_s}A_{\Phi_d} \frac{\partial \omega_{01}}{\partial \Phi_d}\frac{\partial \omega_{01}}{\partial \Phi_s}|\ln (\zeta/ f_{\rm ir}t_m)|\right)^{1/2}
\end{align}
Assuming perfect correlation ($c_{sd}=1$) between the two noise sources \cite{Gus11}, the noise amplitude of the correlated noise is related to the individual noise amplitude as
\begin{align}
A_{\Phi_s \Phi_d}=\sqrt{ A_{\Phi_s}A_{\Phi_d}}.
\end{align}
In the case where there is no correlation between the two noise sources ($S_{\Phi_s\Phi_d}=0$), the dephasing rate becomes
\begin{align}
\Gamma_{\phi} = \sqrt{\Gamma_{\phi,s}^2+\Gamma_{\phi,d}^2}.
\end{align}
\end{appendix}

\end{document}